# Layer compression and enhanced optical properties of few-layer graphene nanosheets induced by ion irradiation

Yang Tan,[†,*] Zhen Shang,[†] Shavkat Akhmadaliev,[‡] Shengqiang Zhou,[‡] and Feng Chen,[†,**]

[†]School of Physics, State Key Laboratory of Crystal Materials, Shandong University, Shandong, Jinan, 250100, China

[‡]Institute of Ion Beam and Materials Research, Helmholtz-Zentrum Dresden-Rossendorf, Dresden 01314, Germany

Graphene has been recognized as an attractive two-dimensional material for fundamental research and wide applications in electronic and photonic devices owing to its unique properties. The technologies to modulate the properties of graphene are of continuous interest to researchers in multidisciplinary areas. Herein, we report on the first experimental observation of the layer-to-layer compression and enhanced optical properties of few-layer graphene nanosheets by applying the irradiation of energetic ion beams. After the irradiation, the space between the graphene layers was reduced, resulting in a tighter contact between the few-layer graphene nanosheet and the surface of the substrate. This processing also enhanced the interaction between the graphene nanosheets and the evanescent-field wave near the surface, thus reinforcing the polarization-dependent light absorption of the graphene layers (with 3-fold polarization extinction ratio increment). Utilizing the ion-irradiated graphene nanosheets as saturable absorbers, the passively Q-switched waveguide lasing with considerably improved performances was achieved, owing to the enhanced interactions between the graphene nanosheets and evanescent field of light. The obtained repetition rate of waveguide laser was up to 2.3 MHz with a pulse duration of 101 ns.

KEYWORDS: Graphene, Ion irradiation, Q-switched waveguide laser, Layer-to-layer compression, ion implantation.

Since the discovery in 2004, graphene has attracted great interest as a monolayer graphite sheet for a broad spectrum of multidisciplinary applications in electronics, optics, biology, and material sciences.[1,2] Graphene has saturable absorption due to the Pauli blocking effect[3] and hot luminescence resulting from nonequilibrium carriers.[4] The unique photonic and electronic properties enable graphene as intriguing candidate for various electronic and optoelectronic devices. For example, graphene has been demonstrated to have the potential to act as a biosensor for DNA and glucose[5] and can be used for the detection of individual gas molecules.[6] In optics, owing to the nonlinear absorption, graphene has been applied for ultrafast laser generation in fibers and bulks as saturable absorbers (SAs) based on the energy-band structure.[7,8] Moreover, graphene nanosheets exhibit strong polarization-dependent absorption of the evanescent field from photonic devices such as waveguides and fibers. This feature enables graphene-coated optical fibers acting as broadband polarizers.[9]

The graphene technology offers large-area few-layer graphene nanosheets of high quality synthesized on metal foils by the chemical vapor deposition (CVD) process.[10,11] To fabricate graphene-based devices, it is usually necessary to transfer the graphene nanosheets from the surface of metals to the target substrates. To achieve a clean, easy transfer process for large-area graphene nanosheets, the transfer technology is required for improvement in terms of speed, uniformity, and close attachment with targets.[12-14] The conventional methods for graphene transfer include the exfoliation of several graphene layers from highly ordered pyrolytic graphite by a mechanical 3M tape, or using electrostatic force and wet-transfer for CVD graphene.[15-17] These processes usually require long treatment cycles with the increment of manufacturing costs, and much effort on the quality control of the transferred graphene.[12]

Recently, ion irradiation has been proved to be an efficient method to provide a perfect lattice of graphene.[18-21] During the ion-irradiation process, the energy of incident ions is transferred onto graphene layers through nuclear collisions and electronic excitations, which are characterized by so-called nuclear ($S_n$) and electronic ($S_e$) stopping powers, respectively.

When $S_n$ or $S_e$ is sufficiently high, controllable defects can be introduced into the few-layer graphene nanosheets.[22-27] Moreover, the ion irradiation of optical materials has been proved to be an efficient method to develop optical waveguide structures by modifying the refractive index of substrates. The successful examples include more than 100 materials of single crystals, ceramics, and glasses.[28-30] With the synergetic ion-irradiation-induced effects from both graphene and optical materials, it is possible to develop novel graphene-on-chip devices for diverse applications.

In this work, we demonstrate the irradiation of energetic ions (e.g., carbon or oxygen) as an efficient method to modify the optical properties of transferred few-layer graphene nanosheets with an area of 1 cm$^2$. The ion irradiation not only reduces the space between the graphene layers, remaining the number of layers unchanged, but enables an enhanced contact between few-layer graphene nanosheets and substrate. Owing to this feature, the polarization-dependent linear absorption of few-layer graphene nanosheets considerably increases. In addition, the ion irradiation produces a planar optical waveguide below the graphene layer. With this one-step irradiation processing, a platform with graphene coated planar waveguide structure has been manufactured. In case of the substrate of laser crystals, such as the Nd:YAG crystal investigated in the work, it is possible to achieve the graphene Q-switched waveguide lasing through the interaction between the evanescent field of waveguide modes and ion-modified few-layer graphene as SA. Excellent performances such as a stable emission, shortened pulse duration, and increased repetition rate are obtained in the passively Q-switched waveguide laser systems. This work also offers a new solution to modify the optical properties of transferred few-layer graphene nanosheets in optical devices by ion irradiation.

The few-layer graphene nanosheets used in this work were fabricated by the CVD method followed by multiple transfers. First, a bilayer graphene was fabricated by the CVD method on copper and nickel disks. Then, the bilayer graphene was transferred four times to prepare few-layer (eight layers) graphene nanosheets. The dimension of the graphene nanosheet was 1 cm$^2$.

Two few-layer (eight layers) graphene nanosheets were irradiated by carbon (Sample 1) and oxygen (Sample 2) ions. The oxyegn ($O^{5+}$)-ion irradiation was carried out at an energy of 15 MeV and a fluence of $2\times10^{14}$ ions/cm$^2$ using a 3 MV Tandetron accelerator at Helmholtz-Zentrum Dresden-Rossendorf, Germany. The carbon ($C^{3+}$)-ion irradiation was carried out at an energy of 6 MeV and a fluence of $1\times10^{15}$ ions/cm$^2$ using a $2\times1.7$ MV tandem accelerator at Peking University, China.

Figures 1a and 1b show the microscopic images of few-layer graphene nanosheets (transferred on a Nd:YAG crystal wafer) obtained by an atomic force microscope (AFM) before and after carbon-ion irradiation (Sample 1), exhibiting an area of 2500 μm$^2$ and a vertical extent of a few nanometers. As shown in Figure 1c, the maximum and minimum values of the surface roughness of the graphene nanosheets before the irradiation were 37 and 7 nm, respectively. The height variation (RMS 6.67 nm) was induced by bubble-like structures, which were formed during the fabrication of few-layer graphene nanosheets through the multiple transfer of the bilayer graphene. A perfect contact between each bilayer graphene is difficult to be achieved over the large-area graphene nanosheets, forming extra space between the adjacent graphene layers. After the carbon ion irradiation, the bubble-like structures inside the few-layer graphene nanosheets disappeared, as shown in Figure 1b, and the height variation of graphene nanosheets (RMS 3.00 nm) was shown as meandering wrinkles. As indicated in Figure 1c, the height of the flat area of the nanosheet was varied from 17 to 5 nm. Extensive analysis of the AFM data shown in Figures 1a and 1b revealed the two variations of graphene nanosheets induced by ion irradiation: (i) the morphology of the graphene nanosheets changed from bubble-like structures to meandering wrinkles (the surface of the graphene nanosheets became smooth); and (ii) the average thickness of the graphene nanosheets decreased from 12.55 nm to 7.66 nm. The few-layer graphene nanosheets possessed an enhanced contact with the Nd:YAG substrate. Two possible reasons may be responsible for this phenomenon. One was that the graphene nanosheets were directly sputtered by energetic ion beams, and the average thickness was reduced due to the decrease in the number of graphene layers. The other was that the number of layers was the same, whilst their appearance was changed by the ion irradiation process.

To confirm the origin of the ion induced effect, we utilized the Raman spectroscopy to investigate the structural and electronic characteristics of the graphene nanosheets with and without ion irradiation. Figure 2a shows the three intense bands in the spectra: (1) the major Raman band of graphene (*G* band), (2) *D* band, and (3) *2D* band. The Raman spectrum shown in Figure 2a was normalized with the maximum intensity of the G band (red line). The intensity of the *D* band clearly increased after the ion irradiation. The intensity ratio of the *D* and *G* bands ($I_D/I_G$) in Raman spectroscopy was used to calculate the zero-dimensional point-like defects in the graphene nanosheets after the ion-beam bombardment. The density of defects was usually determined by the average distance of defects ($L_D$). In the low-defect-density regime, $L_D$ can be calculated as follows[31]:

$$L_D^2(nm^2) = 1.8 \times 10^{-9} \lambda_L^4 (\frac{I_D}{I_G})^{-1} \tag{1}$$

In this work, the $I_D/I_G$ value was increased from 0.26 to 0.65 corresponding to the variation of $L_D$ from 23.5 to 14.9 nm. Clearly, the defect density of graphene nanosheets was increased by the carbon-ion-beam bombardment.

The energy deposition process of the ion bombardment was calculated by the well-known computer program—Stopping and Range of Ions in Matter (SRIM 2013), [32] on the interaction between carbon ions and few-layer graphene/Nd:YAG system. As shown in Figure 2b, the percentage of displaced atoms induced by the ion irradiation was less than 0.01%, which was too low to create significant damage of the graphene layer. This also suggests the maintained number of graphene layers after the ion irradiation.

Further evidence was also obtained by the Raman spectroscopy. The 2D and G bands (shape and intensity) can be used to determine the number of graphene layers.[33] Furthermore, the intensity of the G band exhibited almost a linear relationship with the graphene thickness. Compared to the spectra shown in Figure 2a, the relative intensity ratio of the 2D and G bands was similar for the sample with and without ion irradiation. Thus, it can be concluded that the number of graphene layers did not change after ion irradiation. To conclude, the layer number of the few-layer graphene was not affected by the carbon ion irradiation in this work.

The smooth and compaction of graphene nanosheets can be attributed to the atomic rearrangements during the ion irradiation. During the irradiation, most of the slowing down of the ions occurs due to the loss in electronic energy ($S_e$), particularly in the graphene layer in this case. A large number of near-surface atoms in the graphene layer are set into motion by this electronic excitation. The mobility results in atomic rearrangements including the sputtering and re-deposition of carbon ions, consequently leading to the smoothing and compaction.[34,35] In case of oxygen ion irradiation (Sample 2), the thickness of the graphene nanosheets decreased to 5.17 nm, even less than the carbon ion irradiated sample (i.e., 7.66 nm for Sample 1). Clearly, the space between the graphene layers can be adjusted to a certain value by selecting an appropriate energy of incident ions.

The nonlinear absorption coefficient of the graphene nanosheets was measured by the single-beam Z-scan technology. A 532-nm laser with 4-ns pulse duration and an energy of 1 μJ was focused using a lens (a focal distance of 400 mm), resulting in ~24.5-μm beam waist. A large-aperture lens was used for collecting the transmitted laser light. By moving the graphene nanosheets around the focal point, the power of the transmitted light was measured as a function of the distance from the focal point. The nonlinear absorption coefficient was calculated using the eq 2, where z is the distance of the graphene nanosheet from the focal point, I is the intensity of the focused laser, and $\alpha_0$ is the linear absorption coefficient.

$$\frac{dI}{dz} = -(\alpha_0 + \beta I)I \tag{1}$$

Figures 3a, 3b and 3c show the open-aperture Z-scan curves of Sample 1 (carbon-ion-irradiated graphene nanosheets), Sample 2 (oxygen-ion-irradiated graphene nanosheets), and the original sample (non-irradiated), respectively. The nonlinear absorption coefficients ($\beta$) of the graphene nanosheets were $-2.8 \times 10^{-4}$ and $-3.2 \times 10^{-4}$ m/W for Samples 1 and 2, respectively. Compared with the original sample ($-3.0 \times 10^{-4}$ m/W), the $\beta$ value slightly changed after the irradiation. This indicates that the nonlinear absorption of the graphene nanosheets did not change significantly after ion irradiation.

Graphene nanosheets were prepared by two methods for the linear absorption measurements: 1) Graphene nanosheets were coated on a neodymium-doped yttrium aluminum garnet (Nd:YAG) crystal with dimensions of $10 \times 10 \times 1.5$ mm$^3$. Inside the graphene nanosheets, the Nd:YAG crystal was irradiated by oxygen ions at an energy of 15 MeV and a fluence of $2 \times 10^{14}$ ions/cm$^2$ (Sample 2). The waveguide structure was fabricated on the surface of the Nd:YAG crystal after ion irradiation. 2) The Nd:YAG crystal with same dimensions was irradiated by oxygen ions at an energy of 15 MeV and a fluence of $2 \times 10^{14}$ ions/cm$^2$, and the waveguide structure was fabricated. Then, the graphene nanosheet was coated onto the surface of the Nd:YAG waveguide. The experimental setup is shown in Figure 4a. Linearly polarized light at a wavelength of 1064 nm was coupled to the waveguide using a convex lens. By changing the polarization angle of the input light by the waveplate, the power of the output light was collected using a microscope objective and measured by varying the angle.

The maximum output occurred at 0° or 180°, and the minimum appeared at 90° or 270°. As shown in Figures 3d and 3e, the graphene nanosheets resulted in an extinction of light with different polarization angles. For the graphene without ion irradiation, the absorption of the light at 1064 nm was 1.26 dB ($\alpha_0$) and 0.16 dB ($\alpha_{90}$), respectively. The extinction ratio of the light with two polarizations was 1.1 dB ($\alpha_0 - \alpha_{90}$). Moreover, the extinction ratio of the light with ion irradiation increased to 3 dB. This indicates that the polarization-dependent absorption was enhanced to a certain exntent. Because the graphene nanosheets were contacted tightly onto the surface of the waveguide, a better overlap of the evanescence field and graphene nanosheets occurred.

With the enhanced linear absorption and unchanged nonlinear absorption of the graphene nanosheets, it is possible to apply it for the improvement of the performance in the Q-switched waveguide laser system. The few-layer graphene nanosheets (eight layers) were first transferred onto the surface of a Nd:YAG crystal, and the graphene-coated Nd:YAG crystal was irradiated by oxygen ions at energy of 15 MeV and fluence of $2 \times 10^{14}$ ions/cm$^2$. In this work, the Nd:YAG waveguide structure served as a laser resonator, and the graphene nanosheets were coated as SA. Two specially designed mirrors were adhered to the end-facets of the planar waveguide to construct a Fabry–Perot oscillator. A high reflectivity (HR) at 1064 nm (reflectivity, R >99.9%) was observed with a high transmission (HT) at 810 nm (transmission, T >99.9%) for the input mirror (M1). However, the out-coupling mirror (M2) had a T of ~5% at 1064 nm. A laser with a wavelength of 810 nm, obtained from a Ti:Sapphire CW laser (Coherent MBR PE), was coupled into the waveguide as the pumping laser. The output light was collected using a long-working-distance microscope objective.

The laser emission from the waveguide structure was obtained at a wavelength of 1064 nm as shown in the inset of Figure 5a. The output power of 1064-nm laser as a function of the pumping power is also shown in Figure 5a. The lasing threshold corresponding to the slope efficiency of 2.6% was 75 mW. The maximum output power was 4.9 mW. Typical Q-switched pulse trains were observed with the pumping power over the threshold as shown in Figure 5b. The variations of the pulse duration and repetition rate with the pumping power are shown in Figure 5a. Clearly, the pulse duration was stable at ~101 ns with much higher pumping power than the lasing threshold, a typical phenomenon for a passively Q-switched laser.[36] The repetition rate of the pulsed laser exhibited a linear variation with the pumping power, and the maximum pulse repetition rate was up to 2.3 MHz.

An experiment for lasing was also conducted for a non-irradiated graphene SA as a comparison. The waveguide structure was first produced on the surface of the Nd:YAG crystal (without graphene layer covered) under the same ion-irradiation conditions. Then, the few-layer graphene nanosheets (eight layers) were coated onto the surface of the waveguide. Under the same pumping condition, the laser emission at a wavelength of 1064 nm was also obtained. The output power as a function of the pumping power is shown in Figure 5b. However, no regular pulse train was observed in this experiment. This can be explained by the previous discussion on the linear absorption of graphene as shown in Figures 3d and 3e. Without the ion irradiation of graphene, the contact between the graphene nanosheets and waveguide was not adequate tight. Consequently, the absorption induced by the graphene nanosheets was too low to achieve Q-switching for pulsed lasing. Moreover, a similar design was performed in Ref. [37], in which an index-matching fluid was added onto the surface of the waveguide to increase the absorption of the graphene. However, the absorption of the graphene nanosheets through the evanescent field was still in a low level, generating the pulse duration larger than 10 µs and the repetition rate less than 29 kHz. In this work, with the enhancement of properties of graphene by ion irradiation, the pulse duration was decreased from ~10 µs to ~100 ns, and the repetition rate was 2.3 MHz. This comparison shows advantages of ion-irradiation as an efficient technique to improve the graphene-based Q-switching features.

In conclusion, the ion-irradiation was demonstrated as a novel method to enhance the performance of graphene-based optical devices. For transferred graphene nanosheets on Nd:YAG laser crystal, ion irradiation has shown the capability to improve the smoothness of graphene and to enhance the contact between graphene and Nd:YAG crystal, resulting in enhanced the linear optical absorption of the graphene. By utilizing this technique for passive Q-switching, the pulsed lasing based on the graphene-coated waveguide system was considerably improved, with stable pulses of 101ns duration and 2.3 MHz repetition rate. This work also paves a way to improve performances of graphene-on-chip devices for various applications.

## AUTHOR INFORMATION

Corresponding Author

*E-mail: tanyang@sdu.edu.cn

**E-mail: drfchen@sdu.edu.cn

Notes

The authors declare no competing financial interest.

## ACKNOWLEDGEMENTS

This work is carried out with the financial support by the National Natural Science Foundation of China (No. U1332121). S. Z. acknowledges the financial support from the Helmholtz-Association (VH-NG-713).

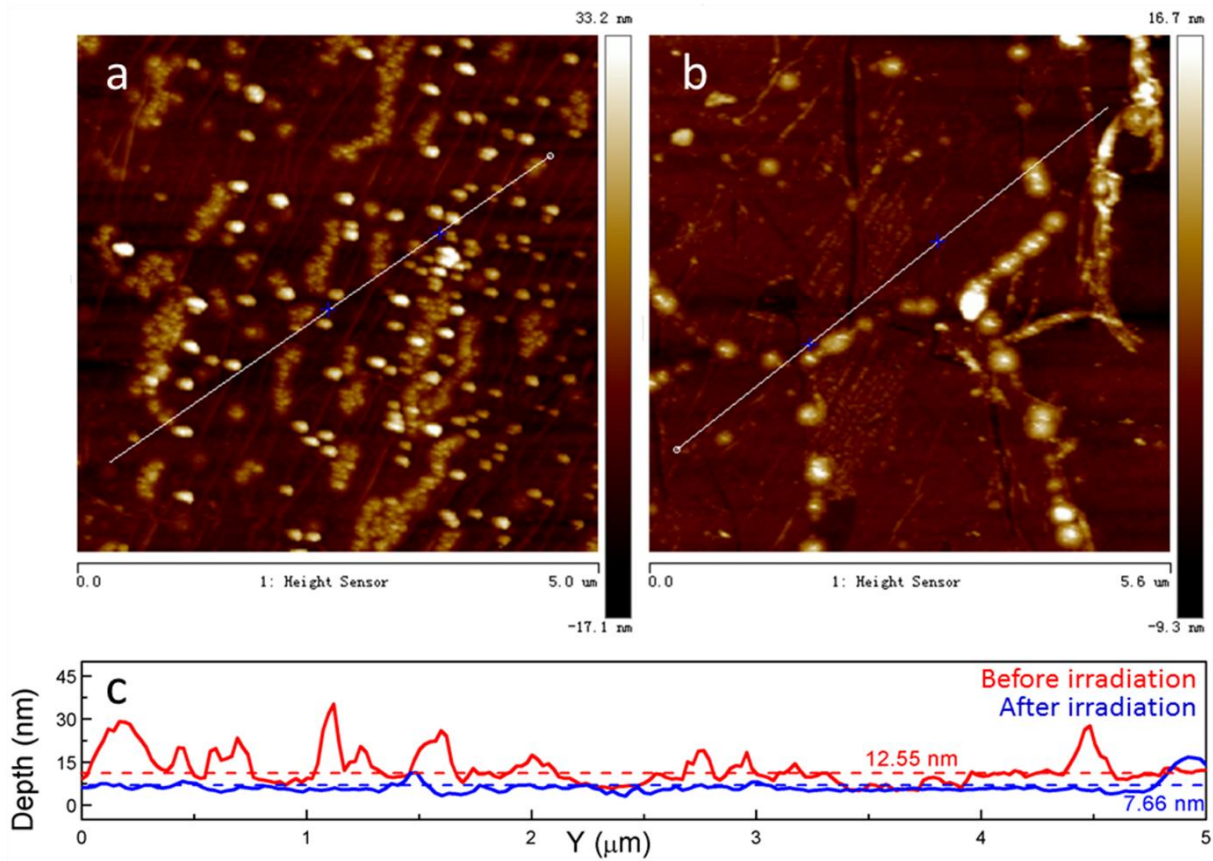

**Figure 1.** Morphology of the few-layer graphene nanosheets before (a) and after (b) ion irradiation captured by AFM. (c) Thickness of the few-layer graphene nanosheets along the lines in (a) and (b).

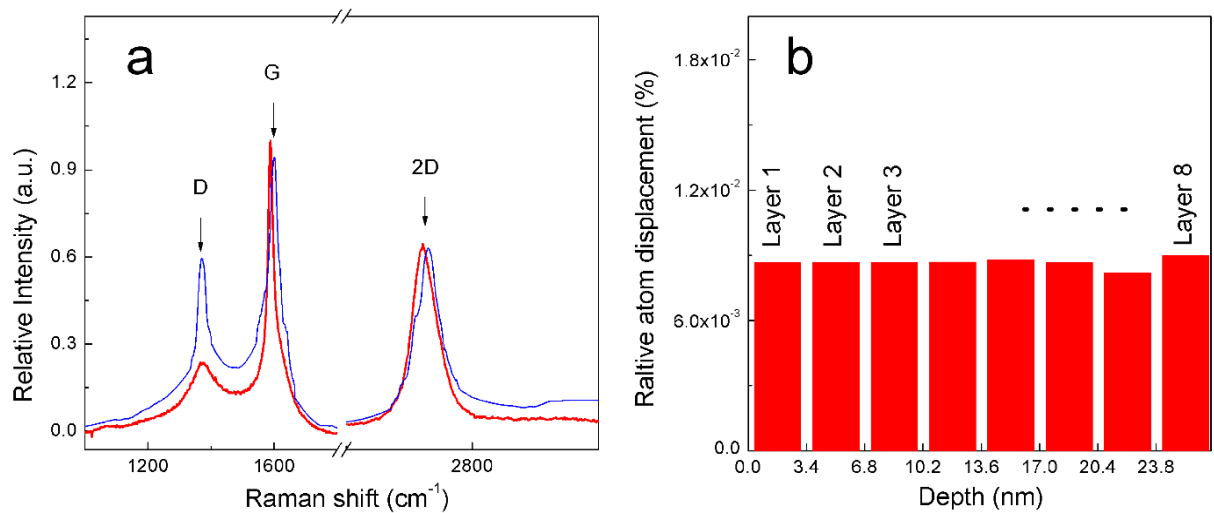

**Figure 2.** (a) Raman spectra of the few-layer graphene nanosheets before and after ion irradiation. (b) Relative atom displacement of the graphene lattice induced by ion irradiation.

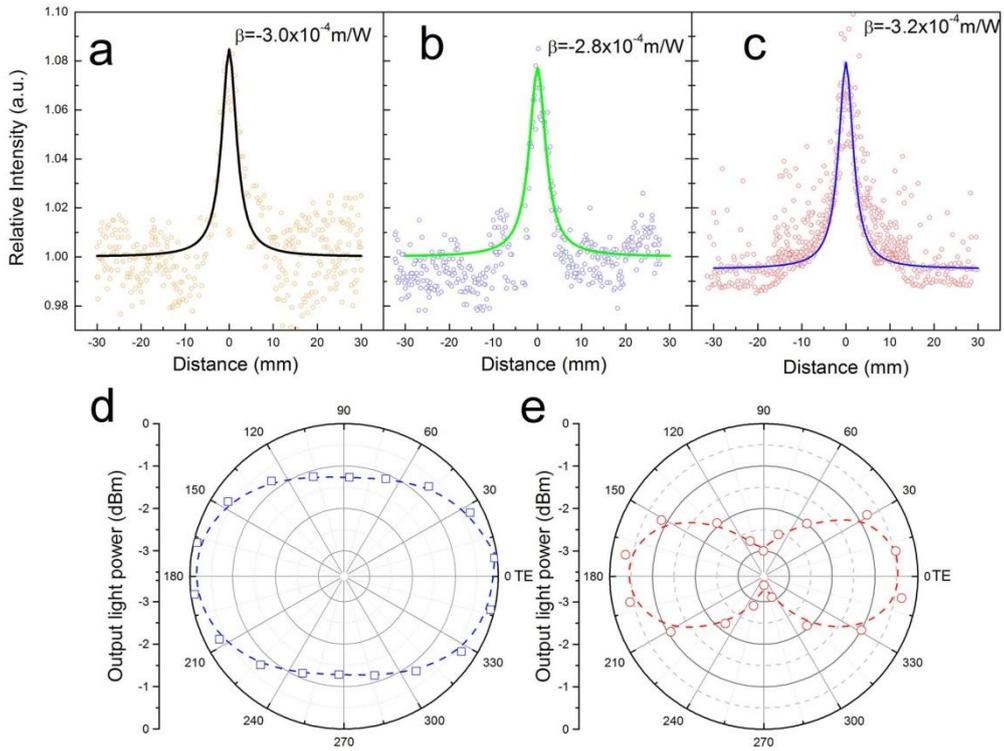

**Figure 3.** Open-aperture Z-scan curves of the original sample (a), Sample 1 (b), and Sample 2 (c). Light extinction of the original sample (d) and Sample 3 (e) in different polarization angles.

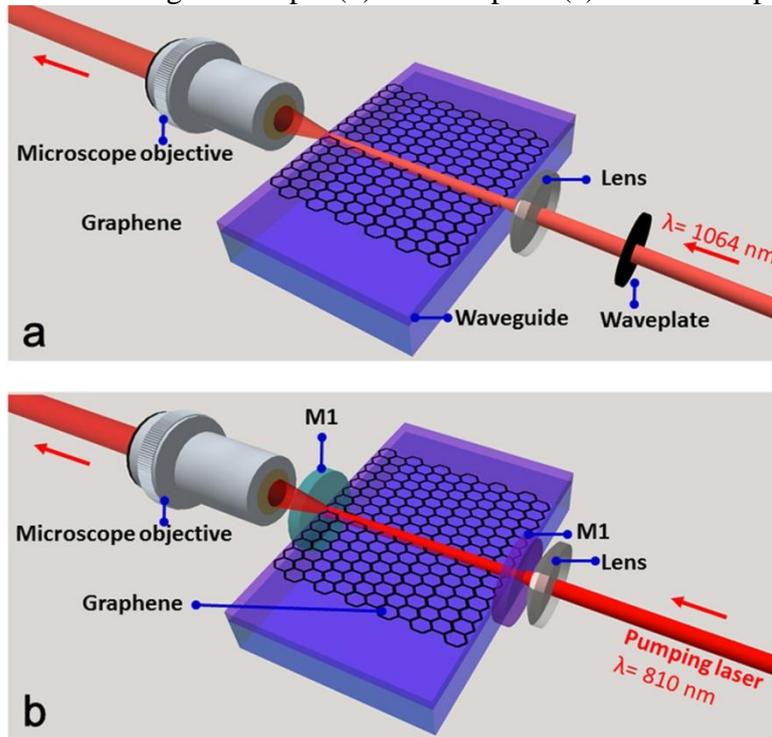

**Figure 4.** (a) Experiment setup for the linear absorption measurements of few-layer graphene nanosheets. (b) Experimental scheme for Q-switched waveguide laser operation.

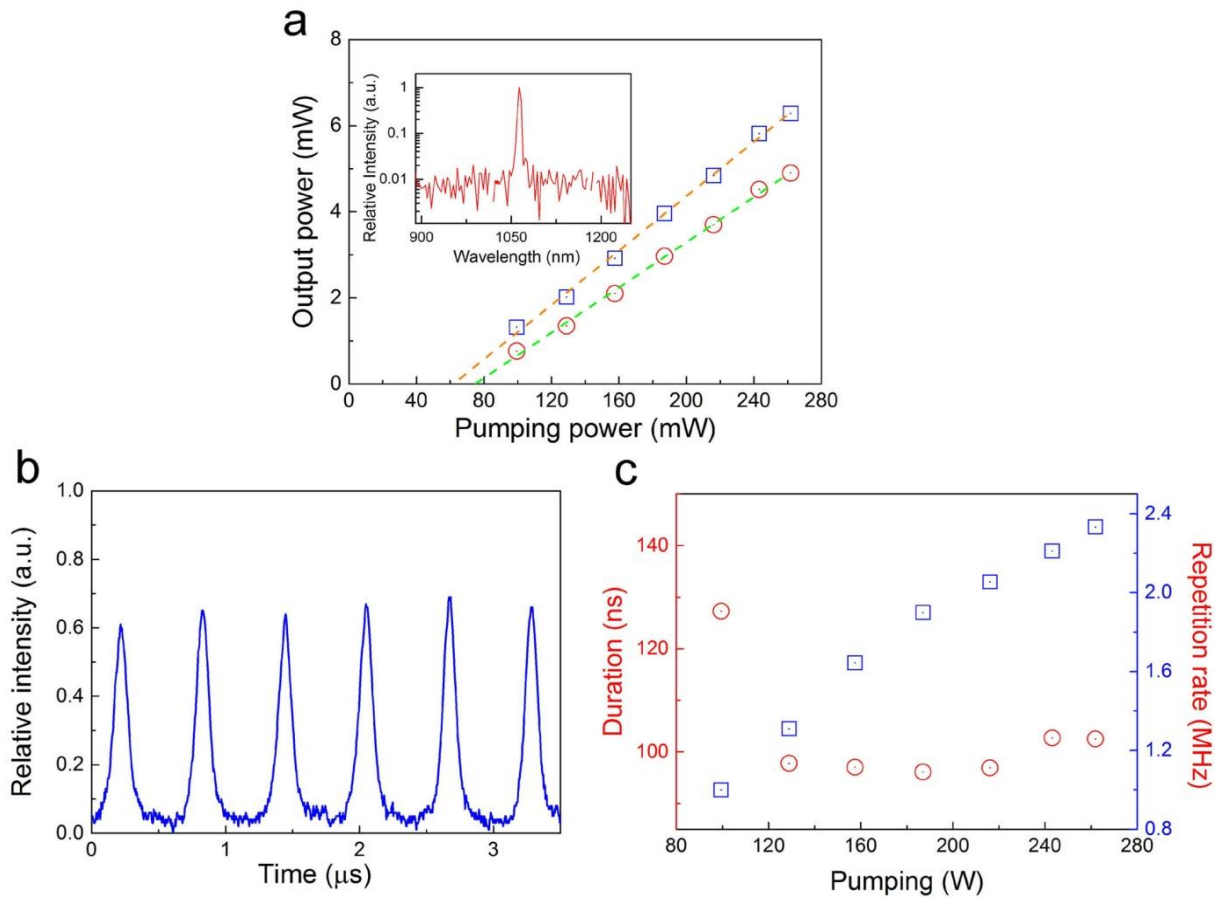

**Figure 5.** (a) Power of the output pulsed laser as a function of the incident pumping power modulated by the few-layer graphene nanosheets without (blue square) and with (red circle) ion irradiation; the inset picture shows the spectrum of the output laser. (b) Pulse train of pulsed waveguide laser with a pumping power of 158 mW. (c) Variation of the pulse duration and the repetition rate of the graphene Q-switched pulsed waveguide laser as a function of the input pump power.

Supporting Information